\documentstyle[12pt]{article}
\catcode`\@=11
\long\def\@makefntext#1{
\protect\noindent \hbox to 3.2pt {\hskip-.9pt
$^{{\ninerm\@thefnmark}}$\hfil}#1\hfill}                

\def\@makefnmark{\hbox to 0pt{$^{\@thefnmark}$\hss}}  

\def\ps@myheadings{\let\@mkboth\@gobbletwo
\def\@oddhead{\hbox{}
\rightmark\hfil\ninerm\thepage}
\def\@oddfoot{}\def\@evenhead{\ninerm\thepage\hfil
\leftmark\hbox{}}\def\@evenfoot{}
\def\sectionmark##1{}\def\subsectionmark##1{}}

\renewcommand{\thefootnote}{\fnsymbol{footnote}}

\newcounter{sectionc}\newcounter{subsectionc}\newcounter{subsubsectionc}
\renewcommand{\section}[1] {\vspace*{0.6cm}\addtocounter{sectionc}{1}
\setcounter{subsectionc}{0}\setcounter{subsubsectionc}{0}\noindent
        {\normalsize\bf\thesectionc. #1}\par\vspace*{0.4cm}}
\renewcommand{\subsection}[1] {\vspace*{0.6cm}\addtocounter{subsectionc}{1}
        \setcounter{subsubsectionc}{0}\noindent
        {\normalsize\it\thesectionc.\thesubsectionc. #1}\par\vspace*{0.4cm}}
\renewcommand{\subsubsection}[1]
{\vspace*{0.6cm}\addtocounter{subsubsectionc}{1}
    \noindent {\normalsize\rm\thesectionc.\thesubsectionc.\thesubsubsectionc.
        #1}\par\vspace*{0.4cm}}

\newcounter{appendixc}
\newcounter{subappendixc}[appendixc]
\newcounter{subsubappendixc}[subappendixc]

\renewcommand{\appendix}[1] {\vspace*{0.6cm}
        \refstepcounter{appendixc}
        \setcounter{figure}{0}
        \setcounter{table}{0}
        \setcounter{equation}{0}
        \renewcommand{\thefigure}{\Alph{appendixc}.\arabic{figure}}
        \renewcommand{\thetable}{\Alph{appendixc}.\arabic{table}}
        \renewcommand{\theappendixc}{\Alph{appendixc}}
        \renewcommand{\theequation}{\Alph{appendixc}.\arabic{equation}}
        \noindent{\bf Appendix \theappendixc #1}\par\vspace*{0.4cm}}

\def\abstracts#1{{
\centering{\begin{minipage}{12.2truecm}\footnotesize\baselineskip=12pt\noindent
        \centerline{\footnotesize ABSTRACT}\vspace*{0.3cm}
        \parindent=0pt #1
        \end{minipage}}\par}}

\renewenvironment{thebibliography}[1]
        {\begin{list}{\arabic{enumi}.}
        {\usecounter{enumi}\setlength{\parsep}{0pt}
\setlength{\leftmargin 1.25cm}{\rightmargin 0pt}
         \setlength{\itemsep}{0pt} \settowidth
        {\labelwidth}{#1.}\sloppy}}{\end{list}}

\newcounter{itemlistc}
\newcounter{romanlistc}
\newcounter{alphlistc}
\newcounter{arabiclistc}

\newcommand{\fcaption}[1]{
        \refstepcounter{figure}
        \setbox\@tempboxa = \hbox{\footnotesize Fig.~\thefigure. #1}
        \ifdim \wd\@tempboxa > 6in
           {\begin{center}
        \parbox{6in}{\footnotesize\baselineskip=12pt Fig.~\thefigure. #1}
            \end{center}}
        \else
             {\begin{center}
             {\footnotesize Fig.~\thefigure. #1}
              \end{center}}
        \fi}

\newcommand{\tcaption}[1]{
        \refstepcounter{table}
        \setbox\@tempboxa = \hbox{\footnotesize Table~\thetable. #1}
        \ifdim \wd\@tempboxa > 6in
           {\begin{center}
        \parbox{6in}{\footnotesize\baselineskip=12pt Table~\thetable. #1}
            \end{center}}
        \else
             {\begin{center}
             {\footnotesize Table~\thetable. #1}
              \end{center}}
        \fi}

\def\@citex[#1]#2{\if@filesw\immediate\write\@auxout
        {\string\citation{#2}}\fi
\def\@citea{}\@cite{\@for\@citeb:=#2\do
        {\@citea\def\@citea{,}\@ifundefined
        {b@\@citeb}{{\bf ?}\@warning
        {Citation `\@citeb' on page \thepage \space undefined}}
        {\csname b@\@citeb\endcsname}}}{#1}}

\newif\if@cghi
\def\cite{\@cghitrue\@ifnextchar [{\@tempswatrue
        \@citex}{\@tempswafalse\@citex[]}}
\def\citelow{\@cghifalse\@ifnextchar [{\@tempswatrue
        \@citex}{\@tempswafalse\@citex[]}}
\def\@cite#1#2{{$\null^{#1}$\if@tempswa\typeout
        {IJCGA warning: optional citation argument
        ignored: `#2'} \fi}}

 1
 1
 1

\font\ninerm=cmr9

\begin{document}

\newcommand{\ieps}{i\epsilon '}
\newcommand{\pslash}[1]{\hbox{/}\kern-.57em {#1}}
\newcommand{\pslashsmall}[1]{\lower.3ex\hbox{/}\kern-.50em {#1}}
\newcommand{\dslash}{\raise.15ex\hbox{/}\kern-.57em\partial}
\newcommand{\lnbp}{\, \ln\left(\frac{1+b}{2}\right)}
\newcommand {\wf}[2]{\psi_{#1}^{\mbox{}^{#2}}}
\newcommand {\wfz}[2]{{\psi_0}_{#1}^{#2}}
\newcommand {\wfb}[2]{{\overline{\psi}}_{#1}^{#2}}
\newcommand {\wfzb}[2]{{{\overline{\psi}_0}}_{#1}^{#2}}
\newcommand {\gv}{\gamma_5}
\newcommand {\prs}{\frac{1+\gamma_5}{2}}
\newcommand {\pls}{\frac{1-\gamma_5}{2}}
\newcommand{\zl}{\left(1+\frac{\delta\lambda}{\lambda}\right)}
\newcommand{\dzl}{\frac{\delta\lambda}{\lambda}}
\newcommand{\zmu}{\left(1-\frac{\delta\mu^2}{\mu^2}\right)}
\newcommand{\dzmu}{\frac{\delta\mu^2}{\mu^2}}
\newcommand{\ztil}[1]{{\widetilde{Z}_{#1}}}
\newcommand{\dztil}[1]{{\delta\widetilde{Z}_{#1}}}
\def\ie{\hbox{\it i.e.}{}}

\vspace{-1cm}
\begin{flushright}
hep-ph/9504321\\
TUM--HEP--219/95\\
April 1995\\
\end{flushright}
\vspace{2mm}

\centerline{\normalsize\bf HIGGS PHYSICS AND THE EQUIVALENCE THEOREM}
\baselineskip=16pt

\vspace*{0.6cm}
\centerline{\footnotesize KURT RIESSELMANN\footnote{Invited
talk given at the Ringberg Workshop on ``Perspectives for electroweak
interactions in $e^+e^-$ collisions'', Tegernsee, Germany (February 5 -- 8,
Munich, 1995).}}
\baselineskip=13pt
\centerline{\footnotesize\it Physik Department,
Technische Universit\"at M\"unchen}
\baselineskip=12pt
\centerline{\footnotesize\it James-Franck Stra\ss e, 85748 Garching, Germany}

\vspace*{0.5cm}
\abstracts{The equivalence theorem is an extremely useful tool
to calculate heavy Higgs {\it and} top-quark effects for processes that have
center-of-mass-energies (much) larger than the $W$ boson mass. After an
explanation of the renormalization procedures involved,
the results for one- and two-loop radiative corrections to the
fermionic Higgs decay, $H\rightarrow f\bar f$, are given and discussed.
Finally, the renormalization scheme dependence is examined, and the reliability
of the perturbative series is investigated.}

\normalsize\baselineskip=15pt
\setcounter{footnote}{0}
\renewcommand{\thefootnote}{\alph{footnote}}

\section{Introduction}
At LEP I and LEP II, heavy Higgs mass effects are suppressed according to
Veltman's screening theorem.~\cite{vel} However, machines like the LHC and
possibly NLC will investigate processes, in which the presence of a Higgs
with sufficiently large mass $M_H$
could cause large nonperturbative effects.
The reason is the proportionality of the Higgs quartic
coupling $\lambda$ to $M_H^2$, which in perturbative treatments leads to
radiative corrections that contain powers of $M_H$ rather than a logarithmic
mass dependence.

It is of interest to study the apparent breakdown of perturbation theory,
and to put
upper limits on the mass of a weakly interacting Higgs boson.  Beyond such an
upper mass limit, the perturbative cross sections for, e.g., LHC processes like
$W^+W^-$ scattering become rapidly unreliable.

The present article describes the systematics of calculating heavy-Higgs-mass
effects by using the Goldstone boson equivalence theorem.~\cite{cororig,cor}
The usual
Lagrangian of the symmetry-breaking sector is used to calculate the
heavy-Higgs-mass corrections in the limit of $M_H\gg M_W$.  Yukawa couplings
are kept without violating the Goldstone theorem.~\cite{goldtheo}
We obtain a
Lagrangian which implements both heavy-Higgs-mass effects {\it and} large
top-quark-mass corrections.  Using a one-loop calculation we show that this
Lagrangian reproduces the full Standard-Model electroweak corrections to the
decay $H\rightarrow t\bar t$ in extremely good approximation. Finally, we
discuss the leading two-loop correction to $H\rightarrow f\bar f$, and we
conclude with remarks on effects due to the use of
different renormalization schemes.\pagebreak

\section{The Goldstone Boson Equivalence Theorem (EQT)}

The equivalence theorem (EQT) is usually discussed in the context of scattering
processes involving the
weak gauge bosons $W^+,W^-,Z$.   We will outline the EQT along these lines,
with the end of this section being devoted to the application of the EQT to the
process $H\rightarrow f\bar f$.

In the case of scattering processes with gauge bosons, the scattering amplitude
of
the longitudinally polarized gauge bosons $W_L^\pm,Z_L$ and the Higgs boson $H$
are enhanced by factors of $M_H^2/M_W^2\propto\lambda /g^2$ relative to
those which involve transversely polarized gauge bosons and the
small electroweak gauge couplings $g$.
Next  we observe~\cite{don} that
in momentum space the longitudinal component
of the vector boson fields is related to the Goldstone boson fields by
\begin{equation}
W^{\pm}_L(k)=\epsilon^{\mu}_L(k)W^{\pm}_{\mu} = w^{\pm}(k) +
{\rm O}\left(\frac{M_W}{k_0}\right),
\end{equation}
where $\epsilon^\mu$ is the polarization vector,
and $k_0$ is the energy component of the four-momentum $k$.

The Goldstone boson equivalence theorem~$^{2,3,6-10}$\
states that
in the limit of a large center-of-mass energy, $\sqrt{s}\gg M_W$, the
scattering amplitudes for $n$ longitudinally polarized vector
bosons $W_L^\pm$,$Z_L$ and any number of other external particles (including
Higgs particles)
are related to the corresponding scattering amplitudes for the
scalar Goldstone bosons $w^\pm$, $z$ (to which $W_L^\pm$, $Z_L$ reduce for
vanishing electroweak gauge couplings $g$) by
\begin{equation}
T(W_L^\pm,Z_L,H,\ldots)=(iC)^nT(w^\pm,z,H,\ldots)+
\mbox{O}(M_W/\sqrt{s}).
\label{eq1}
\end{equation}
The constant $C$ depends on the renormalization scheme used in the
calculation~\cite{bs,he},
\begin{equation}
C=\frac{M_W^0}{M_W}\,\left(\frac{Z_W}{Z_w}\right)^{1/2}
\left[1+{\rm O}(g^2)\right],\label{eq2}
\end{equation}
where the $Z$'s are the wavefunction renormalization constants
for the physical fields $W^\pm$ and the scalar fields $w^\pm$. $C$
is equal to unity for electroweak couplings $g\rightarrow0$ in schemes in which
the renormalization constants are defined at mass scales $m\ll M_H$. We
choose to
renormalize the $w^\pm$, $z$ fields at $p^2=m^2=0$, a choice which
corresponds to massless Goldstone bosons.
Then \cite{bs,he},
\begin{equation}
C=1+{\rm O}(g^2).\label{eq3}
\end{equation}
In the limit of a heavy Higgs boson, $M_H\gg M_W$, the coupling $g$ is much
smaller than the Higgs coupling $\lambda$:
$M_W^2/M_H^2\propto g^2/\lambda\ll 1$, and the gauge couplings can be
neglected. In this approximation, the constant $C$ is equal to unity. Since we
started with the assumption $\sqrt{s}\gg M_W$, we obtain the result
\begin{equation}
T(W_L^\pm,Z_L,H,\ldots)
\begin{array}[b]{c}
    {\ninerm\sqrt{s},M_H\gg M_W}\\
          \approx
\end{array}
T(w^\pm,z,H,\ldots)\, ,
\end{equation}
where the amplitude on the right-hand-side only depends on the quartic Higgs
coup\-ling $\lambda$ and the Yukawa couplings $g_f$, and only involves
scalar and fermion fields.\footnote{
Note that in the limit of zero gauge couplings
the {\it internal} electroweak gauge bosons are also replaced by massless
scalar Goldstone bosons.~\cite{hvel}}
$\,$(We neglect the QCD sector
of the Standard Model throughout this paper.)
This makes the use of the equivalence
theorem an excellent and easy-to-use approxi\-ma\-tion.

The equivalence theorem can also be applied in processes that have no external
electroweak gauge bosons but receive leading radiative corrections through
loops involving $W^+,W^-,Z$, or $H$. Again, the EQT amplitudes will be a good
approximation as long as $\sqrt{s}\gg M_W$ and $M_H\gg M_W$.
E.g., in the case of the decay of
the Higgs particle, the center-of-mass energy is identical to $M_H$.
Therefore, in the limit of $M_H\gg M_W$, the EQT is expected to be an excellent
approximation.

Quantitatively we find that a Higgs mass of $400$ GeV is sufficiently heavy.
This result is based on a comparison of
the one-loop result for the Standard--Model decay $H\rightarrow t\bar t$ based
on our
equivalence-theorem calculation including Yukawa couplings and
the corresponding full electroweak one-loop calculation.
For $M_H>400$ GeV  the two results agree to better than 96\% for $m_t=174$
GeV.

\section{The Lagrangian consistent with the EQT}

All the physics connected with the Higgs particle is determined by the
Lagrangian of the Standard Model, the starting point of our EQT
calculations.
We begin by defining the full Lagrangian for the symmetry-breaking sector of
the Standard Model. It is given by
\begin{equation}
{\cal L}_{SB}=\frac{1}{2}(D_\mu\Phi)^\dagger(D^\mu\Phi)
-\frac{\lambda}{4}(\Phi^\dagger\Phi)^2
+\frac{\mu^2}{2}(\Phi^\dagger\Phi),
\label{eq4}
\end{equation}
where a positive value of $\mu^2$ shifts the minimum of the potential to a
non-zero vacuum expectation value of the field $\Phi$.
The covariant derivative is defined as
$D^\mu = \partial^\mu + ig{\bf W}^\mu \cdot {\bf T}+\frac{1}{2} g' B^\mu$.
The complex field $\Phi$ is written in terms of four real scalar fields,
\begin{equation}
\Phi=
\left(\begin{array}{c@{+}c}
w_1 & iw_2\\
h & iw_3
\end{array}\right).
\label{eq5}
\end{equation}
The gauge couplings $g$ and $g'$ allow for the interaction of the Higgs
sector with the electroweak gauge sector of the Standard Model.
The use of the
Goldstone boson equivalence theorem
corresponds to calculating the physical observables of interest
in the limit of $g,g'\rightarrow0$. This reduces the above Lagrangian
to a SO(4)-symmetric Lagrangian
involving only the scalar fields $h$ and ${\bf w}=(w_1,w_2,w_3)$.

The field $h$ is taken as usual as the component of $\Phi$
which acquires a vacuum expectation value,
$v$.  This spontaneously breaks the
$SO(4)\simeq SU(2)\times SU(2)$ symmetry of the doublet $\Phi$.
By virtue of the Goldstone theorem \cite{goldtheo}, the spontaneous breaking of
the SO(4) symmetry leads to three massless Goldstone bosons.
We write $h$ as $h=v+H$ where $\langle\Omega|H|\Omega\rangle=0$
with respect to the physical vacuum $|\Omega\rangle$.
This allows for a perturbative calculation by expanding the Higgs field
$H$ and the Goldstone boson fields $w_1,w_2,w_3$ around zero field strength.
We obtain the new Lagrangian
\begin{eqnarray}
{\cal L}_H &=&
\frac{1}{2}(\partial_\mu H)^\dagger(\partial^\mu H)
\;+\;\frac{1}{2}(\partial_\mu {\bf w})^\dagger(\partial^\mu {\bf w})
                                                                \nonumber\\
&&-\; \frac{\lambda}{4}\left({\bf w}^4 + 2{\bf w}^2H^2 + H^4\right)
\;-\; \lambda v\left({\bf w}^2H + H^3\right)      \nonumber\\
&&-\;\frac{1}{2}H^2\left(3\lambda v^2 - \mu^2\right)
  \;-\;\frac{1}{2}{\bf w}^2\left(\lambda v^2 - \mu^2\right)\nonumber\\
&&-\; Hv\left(\lambda v^2 - \mu^2\right)\, .
\label{physlagr}
\end{eqnarray}
As usual, the tree level relationship $\mu^2=\lambda v^2$ guarantees
vanishing tadpole contributions {\it and} massless Goldstone bosons.

In addition to the above Lagrangian, the doublet $\Phi$ also interacts with the
left- and right-handed fermion fields,
$\wf{f}{L,R}$, of the Standard Model, with the strength of the interactions
defined by the Yukawa couplings of the theory.  This gives the second
contribution to our EQT Lagrangian.
Choosing the top-bottom-quark generation as example, the Lagrangian ${\cal
L}_F$
governing the interactions between the doublet $\Phi$ and the fermion fields is
\begin{equation}
{\cal L}_F^{(t,b)} \!=\!
- \left( \frac{g_b}{\sqrt{2}}
\left(  \begin{array}{c}\wfb{t}{L}\\ \wfb{b}{L} \end{array} \right)
\Phi\; \wf{b}{R} + {\rm h.c.} \right) 
- \left( \frac{g_t}{\sqrt{2}}
\left(  \begin{array}{c}\wfb{t}{L}\\ \wfb{b}{L} \end{array}\right)
i\tau_2\Phi^\star \wf{t}{R} + {\rm h.c.} \right) .
\end{equation}
Here $g_t,\ g_b$ are the top- and bottom-quark Yukawa couplings, and $\tau_2$
is the complex Pauli
matrix such that $i\tau_2\Phi^\star$ is the charge conjugate of $\Phi$.
The complete EQT Lagrangian is now defined as
\begin{equation}
{\cal L}_{\rm EQT}={\cal L}_{H}+{\cal L}_F.
\end{equation}

For  a zero vacuum expectation value $v$ of the doublet $\Phi$, the presence of
the fermion fields leads to  a $SU(2)\times U(1)$ symmetry of the complete
Lagrangian ${\cal L}_{\rm EQT}$, ($g_b\not=g_t$), or respects the chiral
$SU(2)\times SU(2)$ symmetry ($g_b=g_t$).
The appearance of a non-zero vacuum expectation value
spontaneously breaks the symmetry,
leading to the presence of three massless Goldstone bosons (Goldstone's
theorem) {\it even for non-zero Yukawa couplings}.
\footnote{It is only the inclusion of the
gauge sector that gives masses to the $W$ and $Z$ bosons.
}$\;$
Each fermion receives a
mass that is proportional to the product of its Yukawa coupling and the vacuum
expectation value, $m_f = {g_f v}/{\sqrt{2}}$.
Depending on the value of the Yukawa couplings, the
interaction of the three Goldstone bosons with fermions may either be
SO(3)-symmetric ($g_t=g_b$), or it might be broken
($g_t\not=g_b$).  The latter case features a residual symmetry of the
charge-conjugate fields
$w^+=({w}_1- i\,{w}_2)/\sqrt{2}$ and $w^-=({w}_1+ i\,{w}_2)/\sqrt{2}$.

The Lagrangian ${\cal L}_{F}$ for any other quark
doublet $(q,q')$ of the Standard Model can be obtained by making the
substitution
$(t,b) \leftrightarrow (q,q')$ in the above expression for ${\cal L}_F$.
For a lepton doublet $(\nu ,l)$, one also can
make the substitution $(t,b) \leftrightarrow (\nu ,l)$;  however, one has to
keep in mind that the Standard Model doesn't provide for right-handed neutrino
fields: $\wf{\nu}{R}=0$ for all lepton flavors.

\section{Renormalization of the Lagrangian}

The Lagrangian ${\cal L}_{\rm EQT}={\cal L}_{H}+{\cal L}_F$
provides all the  information necessary to calculate
Green's functions and $\bf S$-matrix elements.
To include quantum corrections,
we start by renormalizing ${\cal L}_H$, introducing all possible counterterms
that respect the unbroken SO(4) symmetry of the Lagrangian, see
Eq.~(\ref{eq4}). This leads to the introduction of the SO(4)-symmetric
wavefunction renormalization $Z_\phi$, and the counterterms
$\delta\lambda$ and $\delta\mu^2$.
These quantities are sufficient to guarantee a finite theory even in the broken
phase, Eq.~(\ref{physlagr}).  In addition, we allow for finite
field renormalization constants, $\ztil{H},\ \ztil{z}$, and
$\ztil{w}$, to properly normalize the  physical
fields of the broken phase (OMS renormalization). The
renormalization can be summarized as
\begin{eqnarray}
&&\lambda \rightarrow \frac{\lambda + \delta\lambda}{Z_\phi ^2},
\phantom{xxxxxxxxx}
\mu^2   \rightarrow \frac{\mu^2 - \delta\mu^2}{Z_\phi}, \\
&&v \rightarrow {Z_\phi}^{1/2}v, \phantom{xxxxxxxxx}\;
H \rightarrow \widetilde{Z}_H^{1/2}{Z_\phi}^{1/2}H ,
\phantom{\frac{\lambda + \delta\lambda}{Z_\phi ^2}}
\label{vren}\\
&&z \rightarrow \widetilde{Z}_z^{1/2}{Z_\phi}^{1/2}z , \phantom{xxxxx}\;
w^\pm \rightarrow \widetilde{Z}_{w}^{1/2}{Z_\phi}^{1/2}w^\pm ,
\label{zfinite}
\end{eqnarray}
The renormalized Lagrangian in terms of the physical fields is
\begin{eqnarray}
{\cal L}_{H,{\rm ren}} &=&
-\;\frac{1}{2}\ztil{H}H^2
\left(3(\lambda +\delta\lambda)v^2 -(\mu^2-\delta\mu^2)\right)
\nonumber\\
&& -\;\frac{1}{2}(\ztil{{w}}{w^+}^2 + \ztil{{w}}{w^-}^2 + \ztil{z}z^2)
\left((\lambda +\delta\lambda)v^2 -(\mu^2-\delta\mu^2)\right)
\nonumber\\
&&-\; \ztil{H}^{1/2}Hv
\left((\lambda +\delta\lambda)v^2 -(\mu^2-\delta\mu^2)\right)
\phantom{\frac{1}{2}}\nonumber\\
&&+\; {\rm interaction\;\; terms}\, ,
\label{physrenlagr}
\end{eqnarray}
The coefficient of the term linear in the field $H$
is fixed as to cancel tadpole contributions to the Higgs one-point function
order by order in perturbation theory.
This fixes $v$ to be the vacuum
expectation value to all orders,
\ie\, $\langle\Omega|H|\Omega\rangle=0$ to all orders.
At tree level, we require $\lambda v^2=\mu^2$.

It should be noted that both $w^\pm$ and $z$ fields have the same mass
coefficient and mass counterterm, and the counterterm structure is identical to
the coefficient of the linear Higgs term. In the presence of Yukawa
interactions the self-energies of the Goldstone bosons
yield $\Pi_z(p^2)\not = \Pi_{w^\pm}(p^2)$ for arbitrary values of $p^2$. It
seems impossible to cancel tadpole contributions and simultaneously keep all
Goldstone fields massless.
However, an explicit calculation~\cite{lytel} shows that
$\Pi_z(0)=\Pi_{w^\pm}(0)=T/v$, where T is the tadpole term. Hence, the OMS
renormalization can be used without
violating the validity of the Goldstone theorem, \ie\,,
$\langle\Omega|H|\Omega\rangle=0$ while
the Goldstone bosons
remain massless at higher orders in perturbation theory.

Next we fix the mass term of the Higgs field. At tree level, using
$\lambda v^2=\mu^2$, we find the Higgs mass to be $M_H^2=2\lambda
v^2$. Conversely, this equation defines the Higgs
coup\-ling $\lambda$ in terms of the physical mass $M_H$ and the physical
vacuum expectation value $v$. At higher orders, the counterterm $\delta\lambda$
is fixed as to preserve this identity,
with the renormalization point at the physical mass value, $p^2=M_H^2$,
rather than $p^2=0$.

Finally, we need to fix the field renormalization constants. In the OMS
renormalization, the propagators of the fields are renormalized as to have unit
residue at the location of the pole.
In the absence of fermion interactions, only one
finite field
renormalization, $\widetilde Z_H$,
is needed to keep the kinetic terms and free propagators in
standard form. \cite{mah} In this case,
the renormalization constant $Z_\phi$ is defined such that the propagators
of the fields
$w^\pm$ and $z$ have unit residue at the location of the pole, and
$\widetilde Z_H$ corrects the Higgs propagator.
Including fermion interactions, we need to
introduce a second finite renormalization constant.
In Eq.~(\ref{zfinite}), we have intentionally introduced the finite Goldstone
boson renormalization constants, $\ztil{z}$ and $\ztil{w}$, in a symmetric way.
However, one of these two quantities is redundant, namely
$\ztil{w}=1$.  This is connected to the fact, that the vacuum expectation
value, renormalized according to Eq.~(\ref{vren}), is related to the muon decay
constant. The decay of the muon, however, is mediated by the $W$ boson rather
than the $Z$ boson.  Hence, both the vacuum expectation value and the fields
$w^\pm$ are renormalized with the same renormalization constant, whereas the
field $z$ obtains an extra finite renormalization in the presence of Yukawa
interactions.

In summary, the counterterms and renormalization constants
contained in the renormalized
Lagrangian ${\cal L}_H$ of Eq.~(\ref{physrenlagr})
have been fixed as to satisfy the following conditions:
(1) $\langle\Omega|H|\Omega\rangle=0$ to all orders, simultaneously fixing the
    pole of the Goldstone boson propagators to be at $p^2=0$;
(2) the real part of the pole of the Higgs propagator is located at its
    physical mass value $M_H$, fixing the quartic Higgs coupling as
    $\lambda=M_H^2/(2v^2)$ to all orders in perturbation theory;
(3) the real parts of the residues of all propagators  are equal to one at the
    pole location.

The expressions for the wavefunction renormalization
constants in terms of the self-energies are:
\begin{eqnarray}
Z_z\; \equiv \;\: \ztil{z}Z_\phi\;  &=&
        1 + \frac{\partial\Pi_z(0)}{\partial p^2}\, ,\\
Z_{w}\;\equiv \;\;\;\;Z_\phi\;\;\; &=&
        1 + \frac{\partial\Pi_{w}(0)}{\partial p^2}
\label{zwdef} \, ,\\
Z_H\; \equiv \; \ztil{H}Z_\phi \; &=&
        1 + \frac{\partial{\rm Re}\Pi_H(M_H^2)}{\partial p^2}\, .
\label{zhdef}
\end{eqnarray}

To illustrate the breaking of the SO(3) symmetry of the Goldstone bosons due to
the presence of Yukawa interactions, we give the explicit expression
for the finite field renormalization $\ztil{z}$  at one loop:~\cite{lytel}
\begin{equation}
\ztil{z} \;\;=\;\;  1 + \frac{\partial\Pi_{z}(0)}{\partial p^2}
- \frac{\partial\Pi_{w^\pm}(0)}{\partial p^2}
\;\;\approx\;\; 1 - \frac{3g_t^2}{32\pi^2}\,,
\label{zfini}
\end{equation}
where all Yukawa couplings except $g_t$  have been
neglected. (Note: for the hypothetical case $g_t=g_b\,(\not=0)$ the result is
$\widetilde Z_z=1$ --- the SO(3) symmetry of the Goldstone bosons would
persist.)

Because  we want to calculate higher order quantum
corrections  including Yukawa interactions, we also need to renormalize the
Lagrangian ${\cal L}_F$.
As in the case of the Lagrangian ${\cal L}_H$, we use multiplicative
field renormalization constants and counterterms for the couplings.
Regarding the Yukawa couplings rather than the fermion masses
as the fundamental parameters of the theory, we need to renormalize the fields
$\wf{f}{L}$ and $\wf{f}{R}$ as well as the coupling $g_f$.
We introduce the replacements
\begin{equation}\phantom{x} g_f\phantom{x}\;\rightarrow\;
\frac{g_f}{Z_\phi^{1/2}}
        ( 1 + \frac{\delta g_f}{g_f} )\,,
\end{equation}
\begin{equation} \wf{f}{R,L}\;\rightarrow\;\;(Z_f^{R,L})^{1/2}\,\wf{f}{R,L}\, .
\end{equation}
In analogy to the OMS renormalization conditions for the Lagrangian ${\cal
L}_H$ of the Higgs sector,
the quantities $Z_f^{R,L}$ and $\delta g_f$ are fixed by requiring that the
fermion propagator has the
real part of its pole equal to the physical mass value, and that the residue of
the propagator at the pole is equal to one.~\cite{boehm}  This concludes the
complete OMS renormalization of ${\cal L}_{\rm EQT}$.

\section{Applications}

The classical use of the equivalence theorem has been the investigation of
vector boson scattering.~\cite{cororig}  The scattering of longitudinally
polarized gauge bosons, $W_L$ and $Z_L$, has been studied by a number of
authors at tree level~\cite{lee} and higher orders.~\cite{unit}

A different application of the equivalence theorem is the calculation of the
leading corrections to the $\rho$ parameter.~\cite{lytel,barb,flei}
This quantity is --- from the
point of view of the EQT --- a low-energy quantity and defined in the gauge
sector of the Standard Model.  However, neglecting the gauge couplings, the
$\rho$ parameter can be written as
\begin{equation}
\rho=\frac{Z_w}{Z_z}=\ztil{z}^{-1},
\end{equation}
with the one-loop result given in Eq.~(\ref{zfini}).

Another example of a ``low-energy'' application is
the two-loop heavy-top-quark contribution to the $Z\rightarrow b\bar b$
coupling which
has been calculated using massless Goldstone bosons~\cite{barb} and arbitrary
values of $M_H$. The validity
of the equivalence theorem was explicitly verified using Ward--Takahashi
identities.~\cite{barb}  The results have been confirmed in an
independent calculation.~\cite{flei}

A different application of the EQT is the decay $H\rightarrow t\bar t$
which we will discuss here in detail.
This decay process features no external gauge bosons.  Yet the EQT is an
excellent tool to calculate the radiative corrections in the couplings $g_t$
and $\lambda$.

\subsection{One-loop electroweak radiative corrections to $H\rightarrow t\bar
t$}
Because of the high mass of the top-quark,
we keep the top-quark Yukawa coupling, $g_t$, but set all other Yukawa
couplings to zero. In this approximation, the Lagrangian ${\cal L}_{\rm EQT}$
is used to calculate the one-loop corrections to $\Gamma\left(H\rightarrow
t\bar t\,\right)$.

The starting point of our analysis is the term of $\cal L_{\rm EQT}$ which
describes the Higgs-fermion interaction. At tree level we have
\begin{equation}
{\cal L}_{\rm Yuk}^f
 = -\,\frac{{g}_f}{\sqrt{2}}\wfb{f}{R}H\,\wf{f}{L} \; + \; h.c. \,.
\label{bornlagr}
\end{equation}
The Born result for the decay width is given by
\begin{equation}
\Gamma _B\left(H\rightarrow f\bar f\,\right)=
{N_c^fM_H\over16\pi}\left(1-{{4{m_f^2}}\over {M_H^2}}\right)^{3/2}\,g_f^2.
\label{eqborn}
\end{equation}
\noindent Here $N_c^f=1$ (3) is the color factor for lepton (quark) flavors.

The renormalized form of Eq.~(\ref{bornlagr})  is
\begin{equation}
{\cal L}_{\rm Yuk}^f
 = -\,\frac{{g}_f(1+\frac{\delta g_f}{g_f})}{\sqrt{2}Z_{w}^{1/2}}
(Z_f^R)^{1/2}\,\wfb{f}{R}\,(Z_H)^{1/2}H\,(Z_f^L)^{1/2}\,\wf{f}{L} \; + \;
h.c. \,.
\label{lyuka}
\end{equation}
Writing $Z_i = 1 + \delta Z_i$  we obtain the
Feynman rule for the $Hf\bar f$ coupling at higher order:
\begin{eqnarray}
&&-i\frac{g_f}{\sqrt{2}}\left( 1
        + \frac{1}{2}\delta Z_H - \frac{1}{2}\delta Z_{w}
        + \frac{1}{2}\delta Z_f^L + \frac{1}{2}\delta Z_f^R
        + \frac{\delta g_f}{g_f} +{\rm O}(\delta^2)\right) \, .
\end{eqnarray}
For the one-loop calculation,
we neglect the terms of ${\rm O}(\delta^2)$.

The radiatively corrected fermionic decay rate
of the Higgs boson can now be calculated using the new Feynman rule for the
Yukawa coupling
and taking into account the one-particle irreducible Feynman diagrams using
$\cal L_{\rm EQT}$. At one loop, there are six triangular diagrams, [internal
lines $(HHt),(ttH),(zzt),(ttz),(wwb)$, and $(bbw)$], which
contribute. The corrected fermionic Higgs decay width is defined by
\begin{equation}
\Gamma\left(H\rightarrow f\bar f\,\right) = (1+\Delta\Gamma)\,
\Gamma _B\left(H\rightarrow f\bar f\,\right)\,.
\label{eqcor}
\end{equation}
The explicit result for $\Delta\Gamma$ is given in Ref.~(17). 
\begin{figure}[tb]
\vspace*{13pt}
\centerline{
\vspace{3.5in}
}
\vspace{0.05in}
\fcaption{Comparison of the one-loop
results for the ratio $1+\Delta\Gamma\equiv\Gamma\left(H\rightarrow
t\bar{t}\right)/
\Gamma_B\left(H\rightarrow t\bar{t}\right)$ obtained
in various approximations with the full one-loop electroweak result
$(g_1,\ g_2,\ g_t,\ g_b \not = 0)$.
The solid curve (EQT) gives the result obtained
using the equivalence
theorem $(g_1,\ g_2 = 0)$  and a nonzero
top-quark Yukawa  coupling $g_t$ corresponding  to
$m_t=174\ {\rm GeV}$. The dot-dashed curve shows the ${\rm O}(\lambda)\hat{=}
{\rm O}\left(G_FM_H^2
\right)$ correction,
and it is equivalent to an EQT curve with $g_t$=0.
}
\label{fig:rc1lp}
\end{figure}
At one loop, $\Delta\Gamma$ consists of terms $\rm
O(\,\lambda\!=\!G_FM_H^2/\sqrt{2}\,)$,  and $\rm
O(\,g_t^2\!=\!\sqrt{2}G_Fm_t^2\,)$.
In Fig.\ \ref{fig:rc1lp},  we show the size of these corrections as a function
of $M_H$, and compare them with the full
one-loop electroweak correction including electroweak gauge-couplings
$g_1, g_2$ as well as
all Yukawa couplings.~\cite{hff}
The full correction was evaluated in the on-shell renormalization
scheme using $m_t=174\ {\rm GeV}$.
We see that the ${\rm O}\left(\lambda\right)$ term underestimates the full
one-loop electroweak correction term by 32\% (24\%) at
$M_H=500 \ {\rm GeV}$ (1~TeV). However, the complete EQT result including the
top-quark Yukawa coupling reproduces the full one-loop electroweak result
very well. The result obtained using the equivalence theorem with $g_t\not=0$
is only 3.9\% (1.8\%) larger than the full
electroweak one-loop term at $M_H=500\ {\rm GeV}$ (1 TeV) for $m_t=174\
{\rm GeV}$. The use of the equivalence theorem therefore gives a quite
accurate approximation to the full theory, even for the rather low
values of $M_H$ with which we are concerned. The small residual differences
away from the decay threshold
can be accounted for by the transverse gauge couplings,
the nonzero masses of the
$W$ and $Z$ bosons, and the finite masses and Yukawa couplings for the
remaining
fermions. The extra structure of the full electroweak correction close to the
threshold, $M_H=2m_t$, is the result of
virtual-photon exchange in QED.
This generates a Coulomb singularity
and a correction that behaves near threshold as
$1+\alpha_{\rm em}Q_t^2[(\pi/2\beta)+{\rm O}(1)]$,
where $Q_t$ and $\beta$ are the  top-quark electric charge and velocity;
see left end of the dashed line in Fig.\ \ref{fig:rc1lp}.\pagebreak

\subsection{Two-loop radiative corrections to $H\rightarrow f\bar f\;$:
${\rm O}\left(\lambda^2\right)$}

We now describe the calculation of the two-loop correction,
$\rm O\left(\lambda^2\right)$.  For $M_H>m_t$ (which always should be satisfied
in the EQT limit $M_H\gg M_W$), it is the dominant correction to the decay of
the Higgs into any fermion pair $f\bar f$.
All subleading two-loop electroweak corrections, those of
${\rm O}(g_f^2\lambda)$
and ${\rm O}(g_f^4)$, are neglected. It should be noted, that the
dominant correction is flavor-independent, whereas the subleading corrections
depend on the fermionic decay channel considered.

Since the dominant correction is independent of $g_f$, we need to identify the
renormalization pieces that are independent of the Yukawa couplings.
Looking at Eq.~(\ref{lyuka}) we find that $Z_H$ and $Z_w$ are the only
quantities that obtain pure Higgs coupling corrections, \ie\ , terms of order
$\rm O\left(\lambda^n\right)$.  All other quantities, including
the Feynman diagrams for the vertex corrections, receive contributions
proportional to $g_f^2$ or higher powers.
Therefore, we obtain the general result for the leading corrections to all
orders in $\lambda$ to be
\begin{equation}
1+\Delta\Gamma(\lambda) = \left.{{Z_H}\over {Z_w}} \right|_{g_f=0}
\end{equation}

The wave-function renormalization constants
$Z_H$ and $Z_w$ were calculated to two loops,
${\rm O}\left(\lambda^2\right)$, in Ref.~(12) 
using dimensional regularization and OMS renormalization.
Calculating $\Delta\Gamma$, the divergent pieces cancel, and
the ${\rm O}\left(\lambda^2\right)$  electroweak corrections to the
fermi\-onic decay rates emerge naturally as~\cite{hfflong}
\begin{equation}
1+\Delta\Gamma = {{Z_H}\over {Z_w}} = { {1 + a_w\hat{\lambda} +
b_w\hat{\lambda}^2}\over    {1 + a_H \hat{\lambda} + b_H \hat{\lambda}^2} }\,.
\label{eqres}
\end{equation}
The coefficients
in the expansion above have been given analytically~\cite{hfflong} and have
been confirmed.~\cite{ghi}
The numerical values are:
\begin{equation}
\label{eqcoeff}
\begin{array}{ll}
a_w\;=\;1\;,           & b_w\;\approx\; 6.098\;, \\
a_H\;\approx\;-1.12\;, & b_H\;\approx\;41.12 \,.
\end{array}
\end{equation}
The one-loop coefficients $a_H$ and $a_w$ are similar in magnitude, but the
two-loop coefficients $b_H$ and $b_w$ differ in magnitude by roughly a
factor of 7, despite the fact that almost the same number of diagrams,
with similar structures and magnitudes, contribute.
It is also interesting that the coefficients in $Z_H^{-1}$ alternate
in sign; those in $Z_w^{-1}$ do not.

The above expression for $\Delta\Gamma$
automatically resums one-particle-reducible Higgs-boson self-energy
diagrams. However, it
is clear that the resummation contains
only limited information on higher-order terms.
Since we actually have no control of terms beyond
${\rm O}\left(\hat{\lambda}^2\right)$, and are not aware of a physical
principle which would select this as an optimum resummation scheme, we
expand Eq.\ (\ref{eqres}) and discard terms beyond
${\rm O}\left(\hat{\lambda}^2\right) = {\rm O}\left(G_F^2M_H^4\right)$.
This gives the alternative representation
\begin{eqnarray}
1+\Delta\Gamma = {{Z_H}\over {Z_w}} &=& 1+(a_w-a_H)\hat{\lambda}
+\left(b_w - b_H - a_wa_H + a_H^2\right)\hat{\lambda}^2 \label{eqexp}\\
&\approx& 1 + 2.12\hat{\lambda} - 32.66\hat{\lambda}^2 \label{eqoms}\\
&\approx&1+11.1\%\left({M_H\over1\,{\rm TeV}}\right)^2
- 8.9\%\left({M_H\over1\,{\rm TeV}}\right)^4\,.
\label{percent}
\end{eqnarray}
The result agrees at ${\rm O}\left(\hat{\lambda}\right)$
with the known one-loop result.~\cite{vel,mar}
\begin{figure}[tb]
\vspace*{13pt}
\centerline{
\vspace{3.5in}
}
\vspace{0.05in}
\fcaption{Complete ${\rm O}\left(\lambda\right)$ and
${\rm O}\left(\lambda^2\right)$ correction factors
for $\Gamma\left(H\rightarrow f\bar f\,\right)$
for $100 \ {\rm GeV}$ $\leq$ $ M_H$ $\leq 1700 \ {\rm GeV}$.
These corrections are universal, i.e., they are independent of the flavor of
the final-state fermions.
In each order, the expanded result given in Eq.\ (\protect\ref{eqexp})
is compared
to the calculation where the one-particle-reducible Higgs-boson
self-energy diagrams are resummed as shown in Eq.\ (\protect\ref{eqres}).
The two-loop correction cancels the one-loop correction
at $M_H=1114 \ {\rm GeV}$ and is twice as large as the latter,
with an opposite sign, at $M_H=1575 \ {\rm GeV}$.
}
\label{fig:rc2lp}
\end{figure}

We are now in a position to explore the phenomenological implications
of our results.
In Fig.\ \ref{fig:rc2lp}, we show the leading electroweak
corrections to $\Gamma\left(H\rightarrow f\bar f\,\right)$ in the one- and
two-loop approximations with and without resummation of
one-particle-reducible higher-order terms plotted as functions of $M_H$.
We will concentrate first on the expanded results given in Eq.\ (\ref{eqexp}).
While the ${\rm O}\left(\lambda\right)$ term
(upper solid line in Fig.\ \ref{fig:rc2lp}) gives
a modest increase of the rates, e.g., by 11\% at $M_H=1 \ {\rm TeV}$,
the situation changes  when the two-loop term is included.
The importance of this term, which grows as $M_H^4$, increases
with $M_H$ in such a way that it cancels the one-loop term completely for
$M_H=1114 \ {\rm GeV}$, and is twice the size of the one-loop term, with the
opposite sign, for $M_H=1575 \ {\rm GeV}$. The total two-loop correction, shown
by the
lower solid line in Fig.\ \ref{fig:rc2lp}, is then negative and has the same
magnitude as the one-loop correction alone. The perturbation series for the
corrections to
$\Gamma\left(H\rightarrow f\bar f\,\right)$ clearly ceases to converge
usefully, if at all, for $M_H\approx 1100\ {\rm GeV}$, or
equivalently, for $\lambda\approx 10$.
A Higgs boson with a mass larger than about 1100 GeV
effectively becomes a strongly
interacting particle.
Conversely, $M_H$ must not exceed approximately 1100~GeV if the standard
electroweak perturbation theory is to be predictive for the decays
$H\rightarrow f\bar{f}$.
Note that one cannot use the usual unitarization
schemes invoked in studies of $W_L^\pm,Z_L,H$ scattering~\cite{lee,uni}
to restore the predictiveness for the heavy-Higgs width, as no
unitarity violation is involved.

One might expect to improve the perturbative result
in the upper range of $M_H$ somewhat
by resumming the one-particle-reducible contributions to the Higgs-boson
wave-function renormalization by using Eq.\ (\ref{eqres}) rather than
Eq.\ (\ref{eqexp}).
This leads to an increase of the one-loop correction
(upper dotted line in Fig.\ \ref{fig:rc2lp}),
while the negative effect of the two-loop correction
is lessened (lower dotted line) for large values of $M_H$.
However, in the mass range below $M_H=1400 \ {\rm GeV}$,
this effect is too small
to change our conclusions concerning the breakdown of perturbation theory.
Moreover, the resummed expression for the one-loop terms in the perturbation
expansion, when reexpanded to ${\rm O}\left(G_F^2M_H^4\right)$,
does not yield a proper estimate
for the size of the two-loop terms. There is consequently no reason to
favor this approach to the present problem.

The subleading two-loop electroweak corrections, those of
${\rm O}\left(g_f^2\lambda\right)$
and ${\rm O}\left(g_f^4\right)$, are
 still unknown, but one may
estimate their likely importance by comparing the
top-quark Yukawa-coupling correction
to the Higgs-coupling correction at one loop.

\section{Scheme-dependence of the ${\rm O}\left(\lambda^2\right)$
radiative corrections: OMS versus $\overline{\rm MS}$ scheme}

So far we have carried out the renormalization of ${\cal L}_{\rm EQT}$ using
OMS. We
found the dominant correction to $H\rightarrow f\bar f$ to be
(Eq.~(\ref{eqoms}))
\begin{eqnarray}
\label{oms}
1\, +\, \Delta\Gamma_{\rm OMS}
 &\approx& 1\, +\, 2.12\hat{\lambda}_{\rm OMS}\, -\, 32.66\hat{\lambda}_{\rm
 OMS}^2\,  +\, {\rm O}\left(\hat{\lambda}_{\rm OMS}^3\right) ,
\end{eqnarray}
where $16\pi^2\hat{\lambda}_{\rm OMS} = M_H^2/(2v^2)$.
It is interesting to check whether the convergence of the perturbative series
can be improved when using $\overline{\rm MS}$ renormalization. Since the
tree-level result of the fermionic Higgs decay, Eq.~(\ref{eqborn}), is
independent of the coupling
$\lambda$,  we only need the one-loop relation between ${\lambda}_{\rm OMS}$
and ${\lambda}_{\overline{\rm MS}}$ to convert the two-loop OMS result into
$\overline{\rm MS}$.  It is~\cite{sirzuc}
\begin{equation}
\label{coup}
\hat{\lambda}_{\overline{\rm MS}}=
\hat{\lambda}_{\rm OMS}\left[1\, +\, \left(25-3\pi\sqrt{3}
+ 12\ln({\mu^2}/{M_H^2})\right)\hat{\lambda}_{\rm OMS}\,
+\, {\rm O}\left(\hat{\lambda}_{\rm OMS}^2\right)
\right],
\end{equation}
where $M_H$ is the physical Higgs mass, and $\mu$ is the mass scale introduced
in dimensional regularization.
We see that the OMS and the $\overline{\rm MS}$ couplings are equal for
$\mu\approx 0.697 M_H$. Combining the two previous equations,~\cite{willey} we
obtain the correction to the fermionic Higgs decay in $\overline{\rm MS}$
quantities:
\begin{eqnarray}
\label{msbar}
1\, +\, \Delta\Gamma_{\overline{\rm MS}}
 &\approx& 1\, +\, 2.12\hat{\lambda}_{\overline{\rm MS}}\,
-\, \left(51.03 - 25.41\ln({M_H^2}/{\mu^2})\right)
\hat{\lambda}_{\overline{\rm MS}}^2\,
+\, {\rm O}\left(\hat{\lambda}_{\overline{\rm MS}}^3\right).
\end{eqnarray}
The OMS correction given in Eq.~(\ref{oms}) and the
$\overline{\rm MS}$ result are also identical for $\mu\approx 0.697 M_H$.

Truncating the series at two
loops  leaves a residual $\mu$ dependence which indicates the significance of
the ${\rm O}\left(\hat{\lambda}_{\overline{\rm MS}}^3\right)$ terms.
In Fig.~\ref{figmscorr} we show the ${\overline{\rm MS}}$ correction as a
function of
$M_H$, keeping $\mu$ or the ratio $\mu/M_H$ fixed at different values.
Choosing $\mu\approx
0.697M_H$ the two-loop OMS result of  Fig.~\ref{fig:rc2lp} is reproduced.

\begin{figure}[tb]
\vspace*{13pt}
\centerline{
\vspace{3.5in}
}
\vspace{0.05in}
\fcaption{The two-loop correction $\Delta\Gamma_{\overline{\rm MS}}$ as a
function of $M_H$.  The curves show the results when either keeping
$\mu$ fixed at the value of $200 GeV$, or keeping the ratio $\mu/M_H$ fixed at
the values indicated. For $\mu\approx0.7M_H$ the two-loop OMS result of
Fig.~\ref{fig:rc2lp} is reproduced.
}
\label{figmscorr}
\end{figure}

It is interesting to note that for fixed $M_H$ a value  $\mu < 0.697 M_H$
improves the convergence of the perturbative $\overline{\rm MS}$ series
twofold: on one hand the value of $\hat{\lambda}_{\overline{\rm MS}}$ decreases
as $\mu$ becomes smaller
(see Eq.~(\ref{coup})), on the other hand the two-loop coefficient
of the ${\overline{\rm MS}}$ correction also decreases in magnitude
for decreasing $\mu$, vanishing for $\mu=0.366 M_H$ (see
Eq.~(\ref{msbar})) .  For values $\mu > 0.697 M_H$ the opposite is
true: the convergence of the series, as indicated by terms up to two
loops, gets worse in a twofold way as $\mu$ increases.
It seems as if the naive choice of $\mu=M_H$ is not necessarily well
motivated.

Varying the scale $\mu$ in the range $M_H/2<\mu<2M_H$ we already find
indications for significant three-loop contributions (needed to reduce the
$\mu$ dependence) for $M_H>650$ GeV. Explicitly, for $\mu=M_H\,(2M_H)$
we find that the two-loop correction $\Delta\Gamma_{\overline{\rm
MS}} $  is in magnitude equal to the OMS result, with the opposite sign,
for values of $M_H=870\, (650)$ GeV.
However, the size of the two-loop correction is still small (about 3--4\%) for
such values of $M_H$.

\section{Summary}

We have reviewed the equivalence theorem and the approximations involved. The
Lagrangian corresponding to the EQT approximations $\sqrt{s},M_H\gg M_W$ was
formulated and renormalized using OMS conditions.  This Lagrangian is the
basis for calculating top-quark and heavy-Higgs corrections to many physical
observables. We have explicitly discussed the calculation of corrections to the
decay $H\rightarrow f\bar f$. At one loop we find that the EQT calculation
approximates the full electroweak correction very well. Calculating the
dominant two-loop corrections we observe the breakdown of perturbation theory
for values of $M_H$ in the TeV-range. However,  already for values of
$M_H> 650$ GeV we find a significant renormalization scheme dependence of the
${\overline{\rm MS}}$ result, indicating the unreliability of the perturbative
result despite the smallness of the two-loop correction.

\section{Acknowledgements}
I would like to thank B.~Kniehl for inviting me to an interesting and
informative Ringberg workshop,
and the staff of the Ringberg castle for its hospitality.
It is also a great pleasure to thank L.~Durand and B.~Kniehl for discussions
and collaboration on parts of this work, and U.~Nierste for useful discussions.


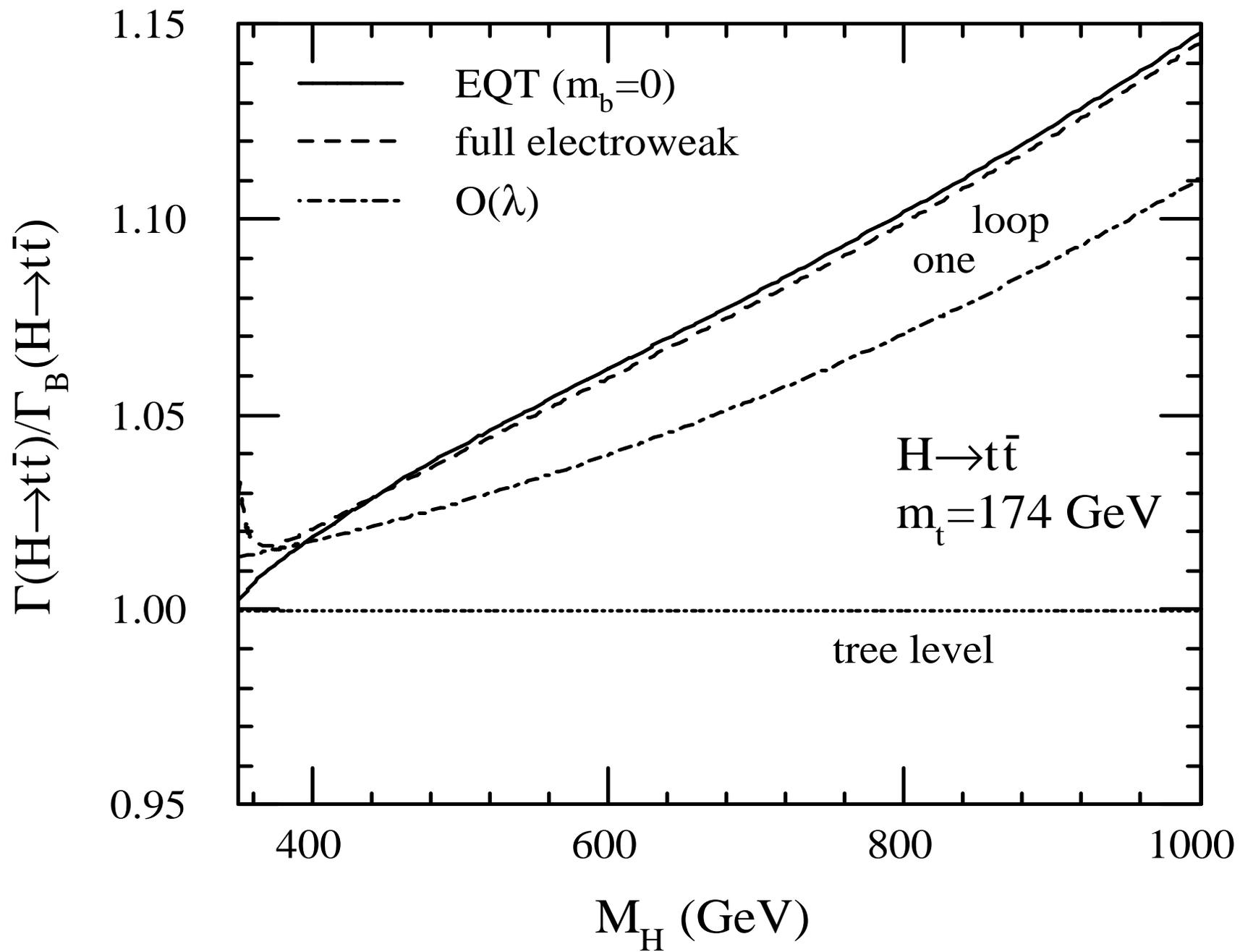
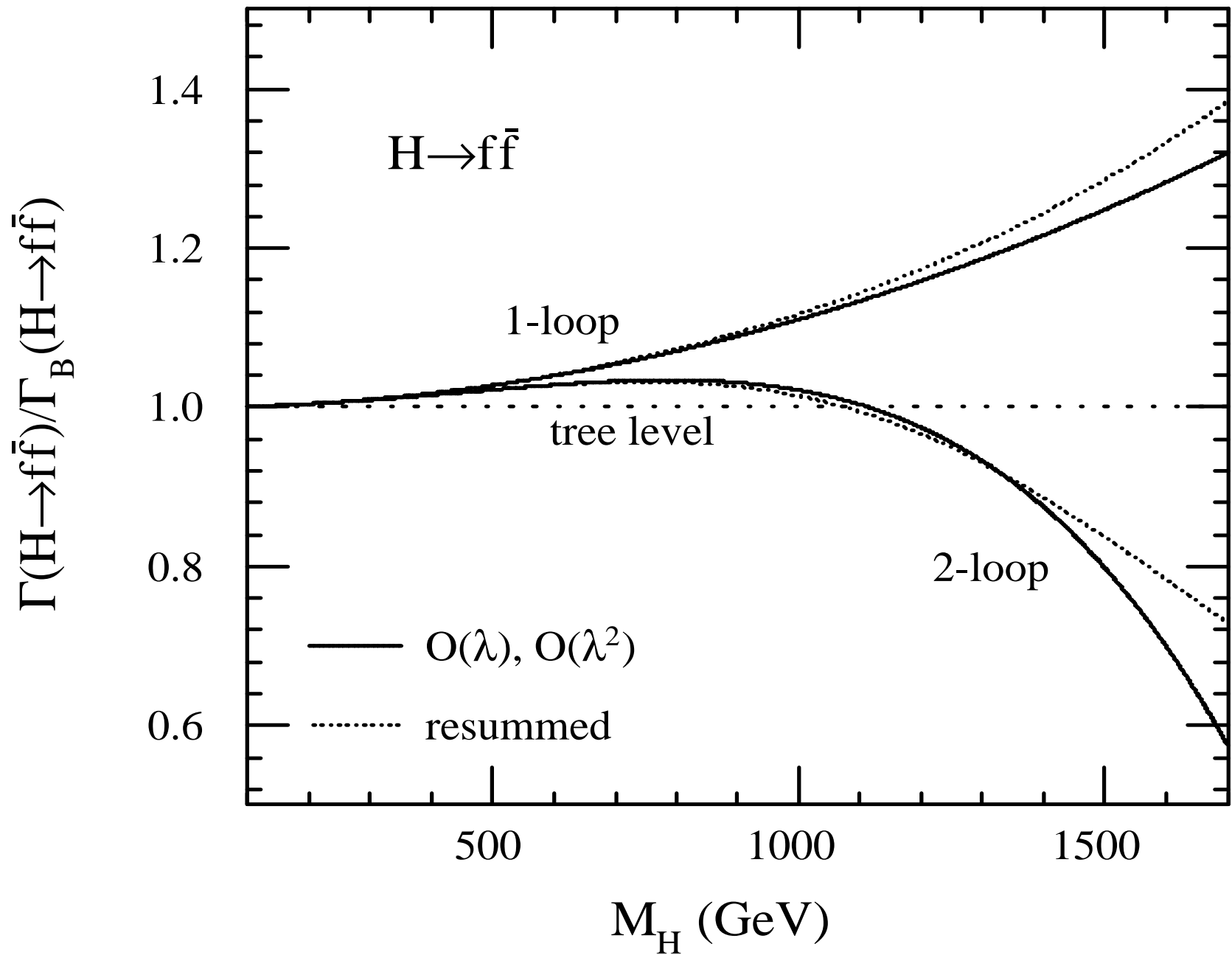
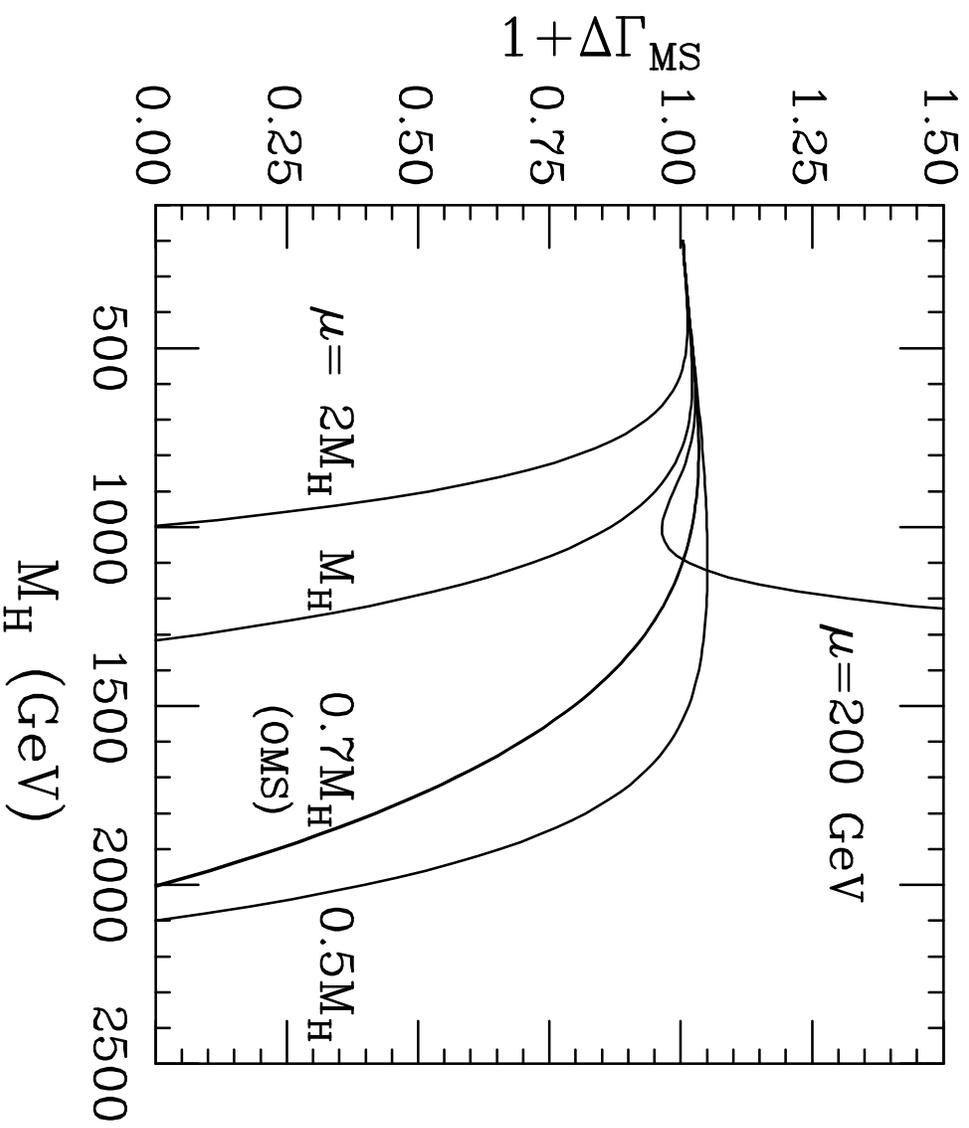
\end{document}